\documentclass[showpacs,showkeys,amsmath,prl,twocolumn]{revtex4}

\usepackage{graphicx}
\usepackage{bm}

\begin{document}

\title{Phase diagram for unzipping DNA with long-range interactions}

\author{Eran A. Mukamel}
 \email{eran@post.harvard.edu}
\author{Eugene I. Shakhnovich}
 \email{eugene@belok.harvard.edu}
\affiliation{Department of Chemistry,
Harvard University, Cambridge, MA 02138}

\date{August 27, 2001}

\begin{abstract}
We present a critique and extension of the mean-field approach to the
mechanical pulling transition in bound polymer systems.  Our model is
motivated by the theoretically and experimentally important examples
of adsorbed polymers and double-stranded DNA, and we focus on the case
in which quenched disorder in the sequence of monomers is
unimportant for the statistical mechanics.  We show how including
excluded volume interactions in the model affects the phase diagram
for the critical pulling force, and we predict a re-entrancy phase
at low temperatures which has not been previously discussed.  We also
consider the case of non-equilibrium pulling, in which
the external force probes the local, rather than the
global structure of the dsDNA or adsorbed polymer.  The dynamics of
the pulling transition in such experiments could illuminate the
polymer's loop structure, which depends on the nature of excluded
volume interactions.
\end{abstract}


\maketitle

\newcommand{\Hpull}{{\cal H_{\text{pull}}}}
\newcommand{\Hpoly}{{\cal H_{\text{0}}}}
\newcommand{\rb}{{\bf r}}
\newcommand{\rxy}{{\bf r^{\|}}}
\newcommand{\Li}{\zeta}
\newcommand{\Zgc}{{\cal Z}}

Recent experiments using micromanipulation techniques to pull and
stretch single molecules have led the way to a better understanding of
statistical properties of polymer systems that cannot be probed in the
bulk~\cite{carlos}.  Theoretical work on these systems has focused on
the complex phenomena arising from heterogeneity in the sequence of
monomers~\cite{nelson}.  In this Letter we study the simpler case of a
homopolymer under the influence of an external force.  We find that
some of the same phenomena caused by disorder in a
heteropolymer can arise in a homopolymer system near its thermal
denaturation temperature due to inhomogeneities in the polymer's
structure.  We also predict a re-entrant phase in the critical force
phase diagram at low temperatures, which should be observable in both
homopolymer and heteropolymer systems.

The Hamiltonian for a polymer in an external field takes the general
form~\cite{doi}
\begin{equation}\label{hpoly}
\Hpoly = \int_0^N \left[\frac{Td}{2b^2} \left(\frac{d\rb}{dn}\right)^2
+ \phi(\rb(n))\right] dn+E[\rb(n)]
\end{equation}
with $d$ the spatial dimension and $b$ the Kuhn length.
Throughout this Letter, we set $k_B=1$.  $\phi(\rb)$ is an attractive
potential arising, for example, from the interaction with an adsorbing
surface, or from the attraction between two strands in double-stranded
DNA (dsDNA).  We
assume that $\phi$ is short-ranged, and depends only on the local
coordinate, $\rb(n)$, describing the distance from the adsorbing
surface or the separation between the two dsDNA strands.  In the case of
a homopolymer, $\phi$ does not depend on the position along the
sequence.  Finally, the volume interactions between different segments
of the chain are represented by the term $E[\rb(n)].$

In the mean-field approximation, interactions between different
sections of the polymer chain, together with whatever external fields
are present, contribute to an overall effective potential, $\phi_{\rm
eff}$, for each monomer.  This field is derived using conditions of
self-consistency, and for low temperatures it should admit bound
states corresponding to adsorbed or zipped conformations.  In this
framework, every monomer far from the ends of the molecule experiences
the same field, and thus contributes the same amount to the overall
bound-state free energy.  The mean-field free energy thus reduces to
${\cal F}_{\rm MF}=N\lambda$, where $\lambda$ is determined by the
self-consistent field, $\phi_{\rm eff}$~\cite{lgk,gros}.  Recent
theoretical work has explored the effect of an external pulling force,
represented by a vector potential term in the energy, on the
mean-field thermodynamics \cite{nelson}.  A polymer that is
adsorbed to an attractive surface or ``zipped'' together with
a complementary chain as in dsDNA, can be pulled out of the bound
state by a force exerted, for example, by a glass bead attached at one
end of the molecule.  The desorbed section of a polymer or the
unzipped part of dsDNA has a force-dependent free energy, ${\cal F}=N
g(F),$ which decreases monotonically with $F$.  The functional form of
$g(F)$ can be obtained from a Gaussian or a freely-jointed chain model
for the stretched polymer.  At a critical force defined by
$g(F_c)=\lambda$, the adsorbed or double-stranded state becomes
unstable with respect to the stretched part of the molecule, and a
sharp unzipping/desporption transition occurs~\cite{nelson}.

The mean-field model relies on the strong assumption that every bound
monomer experiences the same effective field, so that the interaction
between the stretched and bound parts of the molecule is independent
of the number of monomers already pulled out.  This feature of the
mean-field model breaks down, for example, in the presence of
variations in the binding energies for different base pairs in dsDNA,
or sequence disorder in an adsorbed polymer.  The effect of such
sequence heterogeneity on the pulling transition has been the subject
of careful study, and several interesting phenomena have been
predicted to emerge in these systems~\cite{nelson}.  Yet the simple
mean-field model of the pulling transition can break down even for
totally homogeneous polymers when long length-scale correlations in
the structure appear near the temperature of thermal unbinding
(denaturation or desorption), $T_c$.  Intuitively, thermal
fluctuations near $T_c$ cause large loops and sub-chains to form.  The
pulling force interacts with the bound polymer {\em only} in the local
region near the end of the stretched-out part of the molecule, so the
local structure is important in determining the behavior of the pulled
polymer.  Spatial inhomogeneities caused by large loops near $T_c$
affect both the equilibrium and the non-equilibrium behavior of a
bound polymer in response to a pulling force.  We consider these
effects in turn.

The pulling force applied to one end of the polymer adds a vector
potential term in the overall Hamiltonian~\cite{nelson}, giving ${\cal
H}=\Hpoly+\Hpull$, where
\begin{equation}
\Hpull = -{\bf F}\cdot \rb(0) = F \int_{n=0}^N
\left(\frac{dz}{dn}\right)dn- F z(N).
\end{equation}
We have assumed that the pulling force, ${\bf F}=F\hat{z}$, is
perpendicular to the adsorbing surface.  Neglecting the effect of the
finite size of the molecule, we can drop the last term.  After tracing
over all configurations, $\{\rb(n)\}$, in the partition function, we
are left with the following contribution to the free energy from the
pulled-away part of the molecule \cite{nelson}: $g(F) = -\frac{F^2
b^2}{2dT}.$ Alternatively, the freely-jointed chain model, which is
more realistic at high pulling forces and low temperatures, gives:
$g(F) =-T\ln\left[\frac{T \sinh(Fb/T)}{Fb}\right].$

In the mean-field framework, the pulling transition occurs at
$g(F_c)=\lambda,$ or
\begin{equation}
\begin{aligned}
F_c=&\sqrt{2d\lambda T/b^2} \\
F_c=&(T/b)\sinh(F_c b/T) e^{-\lambda /T}
\end{aligned}
 \qquad
\begin{gathered}
\text{(Gaussian)}\\ \text{(Freely-Jointed)}
\end{gathered}
\end{equation}
The mean-field approximation should hold good at low temperatures,
since most monomers will be in the bound state and only short loops
will form.  As $T\rightarrow 0$, the free energy per monomer,
$\lambda,$ converges to some constant value, $-\lambda_0$,
corresponding to the binding energy of one monomer.  The Gaussian
model thus predicts $F_c\sim(\lambda_0 T)^{1/2}$; however, in this
regime the Gaussian model for the free energy of a stretched chain is
not valid, since there is little elasticity in the polymeric bonds.
We instead rely on the freely-jointed model, which gives at low
temperatures: $g(F)\simeq -Fb-T\ln (T/2Fb)$, so that
$F_c(T=0)=\lambda_0/b$.  The limiting behavior for $T\ll\lambda_0 b$
is given by
\begin{equation}\label{lowtemp}
F_c\approx\lambda_0/b-\frac{T}{b}\ln\frac{T}{2\lambda_0b}.
\end{equation}

We thus expect to observe {\em two} desorbed phases, one at high
temperature and another at sufficiently low temperature, for
$F>\lambda_0/b$.  At high temperatures, the entropic advantage of the
unbound monomers overrides the energetic stability of the bound phase.
At low temperatures the free energy per monomer of the stretched part
of the polymer or DNA increases as the chain becomes stiffer.  The
effect of lowering temperature near $T=0$ is to stiffen the chain,
making it easier to pull out the polymer or unzip the dsDNA.  The
temperature at which this re-entrant phase should exist is determined
by the specific properties of the chain, in particular the Kuhn
length, $b$.  In fact, at low temperatures, we expect
$F_c(T,b)=\frac{1}{b}F_c(T,b=1).$ This prediction may allow for
experimental determination of the relative chain stiffness, related to
$b$, for different molecules, through careful low-temperature
investigation of $F_c$.

Close to the thermal unbinding temperature, $T_c$, as $\lambda$
approaches zero, the mean-field approximation breaks down because of
long-range correlations (corresponding to long loops) in the structure
of the bound chain.  Indeed, within the mean-field framework the mean
size of loops grows near $T_c$ as~\cite{gros}: $\langle k\rangle\sim
\lambda^{-1}\sim|T-T_c|^{-2}$.  Forming a loop carries an entropic
cost, which depends non-linearly on the loop size, not simply on the
local density of monomers.  Dealing directly with the interaction
energy, $E[\rb(n)]$, complicates the analytic problem, although
attempts have been made to simulate this full model numerically
\cite{causo}.  We will follow a different route first proposed by
Poland and Scheraga~\cite{poland,kafri}, in which the mean-field model
is extended by introducing a general form for the entropy of a loop of
length $k$,
\begin{equation}
S(k)=A+k\ln(s)-c\ln(k),
\end{equation}
where $s$ is a non-universal constant, and $c$ is determined by the
properties of the loops.  In particular, for non-interacting monomers
($E[\rb(n)]\equiv 0$) the loops are random walks in $d$ dimensions,
with $c=d/2$, whereas excluded volume interactions between monomers
tend to increase the value of $c$~\cite{kafri}.

The binding energy due to the potential, $\phi$, leads to a
statistical weight $\omega= e^{-\phi_0/T}$ for each bounded
base-pair or adsorbed link.  In fact, the only relevant parameter in
this model is the energy difference between the bounded and unbounded
links, $\ln(\omega)-\ln(s),$ so that we can simply put $s=1$.

We work with the grand partition function: $\Zgc(z)=\sum_{m=1}^\infty
Z(m) z^m$, where $Z(m)$ is the canonical partition function for a
chain of $m$ links, and $z$ is a fugacity which controls the size of
the polymer.  In this ensemble, the chain can be regarded as an
infinite collection of loop-train pairs, giving:
\begin{equation}\label{dna1}
\Zgc(z)=\frac{V_0(z)U_L(z)}{1-V(z)U(z)}
\end{equation}
where $U(z)=\sum_{k=1}^\infty \frac{s^k}{k^c}z^k$ and
$V(z)=\sum_{k=1}^\infty \omega^k z^k$. The boundary conditions are
$V_0=1+V(z)$ and $U_L=1+U(z)$.  We will be interested in the
thermodynamic limit, where the total number of links, $\langle
N\rangle=\frac{d\ln\Zgc}{d\ln z}$, diverges.  From Eq.~(\ref{dna1}) we
see that this limit corresponds to the value $z=z^*$, where
$U(z^*)=1/V(z^*) = 1/\omega z^* -1$.  The thermal desorption or
denaturation transition at $T_c$ corresponds to
$z=z_c=1/s$~\cite{poland,kafri}.

We can introduce an external pulling force through an extra boundary
condition: $\Zgc_F=V_F(z)\Zgc,$ where $V_F(z)=1+\sum_{k=1}^\infty
\omega_F^k z^k$, and $\omega_F=e^{-g(F)/T}$ is the statistical weight
per monomer of the stretched part of the polymer.  The number of
monomers pulled away goes like $\langle m\rangle=\frac{d\ln
\Zgc_F}{d\ln\omega_F}\sim(1-\omega_F z^*)^{-1}$, so that we naturally
identify the point $\omega_{F_c}=1/z^*$ as the mechanical pulling
transition.  For $T$ close to $T_c$ this gives,
\begin{equation}
\begin{aligned}
F_c&\sim (T/b)\sqrt{-2d\ln z^*}\qquad\text{(Gaussian)}\\
F_c&\sim (T/b)\sqrt{1-z^*}\qquad\text{(Freely-Jointed)}
\end{aligned}
\end{equation}
At the thermal desorption transition, $z^*\rightarrow z_c=1/s=1$, so
that $F_c\rightarrow 0$.

\begin{figure}
\includegraphics[width=8.5cm]{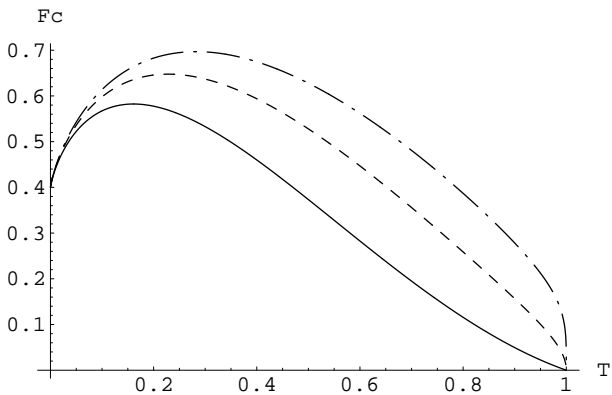}
\caption{\label{phase}
Phase diagram of the critical pulling force,
$F_c$.  The force is given in arbitrary units, and
we have set $\lambda_0=0.4$.  The curves correspond to $c=1.25$ (solid), $c=3/2$ (dashed) and $c\geq 2$ (dot-dashed).  The area below each curve is the bound
phase, while the regime above the curve corresponds to desorbed or
unzipped conformations.}
\end{figure}

The approach to $T_c$ is described by the fraction of bounded
segments~\cite{kafri}, $\theta=\frac{d\ln z^*}{d\ln\omega}$. We
use the fact that, near $T_c$ and with $c>1$, we have
$|U(z_c)-U(z^*)|\sim |z^*-z_c|^\zeta$, where
$\zeta=\min(1,c-1)$.  Thus we obtain $|z^*-z_c|\sim |T-T_c|^{1/\zeta}$, so
that
\begin{equation}\label{mainresult}
F_c\sim |T-T_c|^{\frac{1}{2\zeta}}
\end{equation}
for both the Gaussian and the freely-jointed models.  We illustrate
this result, together with the low-temperature behavior from
Eq.~(\ref{lowtemp}), in Fig.~\ref{phase}.

Note that at $c=3/2$, corresponding to the entropy of non-interacting
loops, Eq.~(\ref{mainresult}) gives $F_c\sim|T-T_c|$, in agreement
with the mean-field result~\cite{gros}, $F_c\sim\sqrt{\lambda}$.  For
$c\geq 3/2, F_c$ has a kink at $T_c$, while for $c<3/2, F_c$
approaches zero with a continuous derivative.  Thus the character of
the mechanical desorption transition changes at $c=3/2,$ in contrast
with the change in the thermal desorption transition, with $\theta$ as
its order parameter, at $c=2$.  Note that the low-temperature behavior
is unaffected by the exponent, $c$, since loop entropy plays no r{\^o}le
in this regime.

Although the value of $c$ affects the order of the thermal
denaturation transition~\cite{kafri}, the mechanical desorption
transition at $F_c$ is always first-order in the case of a
homopolymer.  For low pulling forces, $F<F_c$, there is a free-energy
advantage to keeping as many links as possible bound, while for
$F>F_c$ the molecule would like to unbind as many links as possible.
The desorbed section of the chain is regarded, in the grand canonical
ensemble, as a separate system described by the partition function,
$V_F(z^*)$.  Thus we recover the mean-field results~\cite{nelson}:
$\langle m\rangle\sim |F-F_c|^{-1},$ and $\langle (\Delta
m)^2\rangle\sim |F-F_c|^{-2}$.

Our phase diagram, Fig.~\ref{phase}, agrees qualitatively with very
recent results for pulled polymers modelled as self-avoiding walks
on a hyper-cubic lattice in $d$ dimensions~\cite{maritan}.
Eq.~(\ref{mainresult}) suggests an empirical test for the
exponent, $c$, based on a precise measurement of the behavior of the
critical force near $T_c$.  In particular, experiments on the pulling
of DNA and of adsorbed homopolymers are expected to display different
scaling of $F_c$ near $T_c$.

So far, we have considered the ensemble where the external parameters,
$F$ and $T$, are fixed, so that the pulled polymer remains in
equilibrium during the experiment.  In fact, as long as these
parameters are varied sufficiently slowly, the system will adjust to
changes ``adiabatically.''  Thus, for example, the parameter $m$ will
be characterized by the force-dependent distribution described above,
while the globular part of the polymer will display a corresponding,
temperature-dependent distribution of loop sizes.  The behavior of
this distribution will depend on $c$ close to $T_c$.

Fast variations in the pulling force, on the other hand, may lead to
non-equilibrium behavior, in which the distribution of loop sizes
plays a crucial r{\^o}le.  We can estimate the relevant timescale for this
``non-adiabatic'' pulling from the Rouse model for the dynamics of an
ideal Gaussian chain.  The relaxation time for correlations between
the ends of a chain of length $N$ is $t_N=N^2 b^2 \xi/(3\pi^2 T)$,
where $\xi$ is the coefficient of friction for monomers in the solvent
\cite{doi}.  In general, the relaxation time for a section of the
chain of length $k$ scales as $t_k\sim k^2.$ If the external force
dislodges monomers at a rate of $1/t_F$, we see that for $t_F\gg t_N$
the polymer will remain in equilibrium during the pull-out.

\begin{figure}
\includegraphics[width=8.5cm]{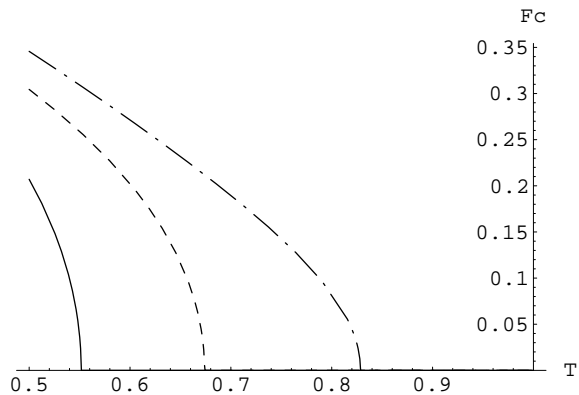}
\caption{\label{phasefast} The critical  force, $\tilde{F_c}$, as a function of
temperature near $T_c=1$, in the non-adiabatic regime, where the pulling timescale
is less than the relaxation time of the polymer loops. The curves
correspond $c=1.25$ (solid), $c=3/2$ (dashed) and $c\geq 2$ (dot-dashed).  Note that in the non-adiabatic regime, $\tilde{F_c}$ vanishes below $T_c$.  The form of these curves depends on the value of $\lambda_0$, the binding energy of a single link, which is here chosen arbitrarily.}
\end{figure}

On the other hand, if $t_F$ is much smaller than the relaxation time
of the smallest loops, the polymer will not have time to re-arrange
itself as $F$ is ramped up, and whole loops will be pulled out at
once.  In this case, the force-extension curve will display jumps
corresponding to the pullout of individual loops.  This behavior is
similar to that derived in models for polymers with sequence
disorder~\cite{nelson}, but the randomness in this case arises from
the distribution of loop sizes.  The loops serve as a source of
quenched disorder in this regime, since they do not have time to
adjust to changes in the mechanical force.
Alternatively, the distribution of loop sizes could become important
even for ``adiabatic'' pulling if the length of the stretched part,
$\langle m\rangle,$ is less than the typical size of loops, $\langle
k\rangle$.  For $c\leq 2$,
we expect the correlation length, and hence $\langle k\rangle$, to
diverge near $T_c$, so that for fixed
$F$ the loop size will eventually grow larger than $\langle m\rangle$,
with a crossover at $\langle m\rangle\approx\langle k\rangle$.

Thus for either a fast-acting external force, or for $T$ sufficiently
close to $T_c$, the pulling force interacts with the polymer
locally, at the level of an individual loop, rather than with the
entire molecule.  Near $T_c$ most of the monomers are unbound, and we
can assume the last loop is attached to the surface or the
complementary strand with only one link.  If the loop contains $k$
unbound links,
its free energy per link is simply $\eta_k=\frac{\ln\omega +S(k)}{k}$.  In this
regime we can write a ``semi-microscopic'' model for the free energy
as a function of the number of pulled-out monomers~\cite{nelson}:
\begin{equation}\label{langevin}
{\cal F}(m)=m g(F)+\sum_{i=1}^{N-m} \eta_k(i)
\end{equation}
where $\eta_k(i)$ is the free energy per monomer of the loop
containing link $i$.  The distribution of loop lengths is simply
$P(k)\propto e^{k\eta_k}=\frac{(sz)^k}{k^c}$. Eq.~(\ref{langevin})
resembles the Langevin equation recently proposed as a model for
the pulling transition in a heteropolymer with sequence
randomness~\cite{nelson}, but in this case the randomness due to
the loop distribution has a non-Gaussian distribution, and thus
cannot be analyzed in the simple framework of Brownian motion.

Eq.~(\ref{langevin}) suggests that the critical pulling force in
the non-adiabatic regime, $\tilde{F_c}$, is smaller than in the
adiabatic regime. This is because the configurational entropy of a
single loop is smaller than that of an ensemble of loops with
variable lengths.  In particular,
\begin{equation}
\tilde{F_c}=\sqrt{-\frac{2dT}{b^2}\langle\eta_k\rangle}
\end{equation}
where the brackets indicate averaging over the distribution of loop
sizes, $\langle\eta_k\rangle\propto\sum_{k=1}^\infty
\frac{\ln\omega+S(k)}{k}\frac{(sz)^k}{k^c}.$  
Fig.~\ref{phasefast}
shows the phase diagram in the non-adiabatic regime near $T_c$, for three values of $c$.  In all three cases, $\tilde{F_c}$ vanishes at a temperature less than $T_c$, in contrast with the adiabatic case.  This temperature is determined by the binding energy, $\lambda_0=-T\ln\omega$, of the single bond at the loop's end.

Note that, as the mean loop size diverges for $c\leq 2$, the
timescale defining adiabatic pulling diverges.  In this regime
even relatively slow variations in the pulling force could be
faster than the relaxation time of the loops.  Simulations and
analytical calculations on the equivalent of Eq.~(\ref{langevin})
for heteropolymers have shown how the size of jumps and plateaux
in the force-extension curve vary as $F$ approaches
$F_c$~\cite{nelson}.  As mentioned above, analysis of
Eq.~(\ref{langevin}) is complicated by the non-Gaussian
distribution of the ``noise'' term, $\eta_k$, but simulations
could elucidate the analogy between the effects of structural
inhomogeneities in pulled homopolymers and sequence disorder in
heteropolymer systems.

The authors would like to thank D. Mukamel and D. Lubensky for important comments on an earlier version of this paper.

\end{document}